# Empirical study and modeling of human behaviour dynamics of comments on Blog posts [*]

Jin-Li Guo

(Business School, University of Shanghai for Science and Technology, Shanghai 200093, China)

**Abstract** On-line communities offer a great opportunity to investigate human dynamics, because much information about individuals is registered in databases. In this paper, based on data statistics of online comments on Blog posts, we first present an empirical study of a comment arrival-time interval distribution. We find that people interested in some subjects gradually disappear and the interval distribution is a power law. According to this feature, we propose a model with gradually decaying interest. We give a rigorous analysis on the model by non-homogeneous Poisson processes and obtain an analytic expression of the interval distribution. Our analysis indicates that the time interval between two consecutive events follows the power-law distribution with a tunable exponent, which can be controlled by the model parameters and is in interval $(1,+\infty)$. The analytical result agrees with the empirical results well, obeying an approximately power-law form. Our model provides a theoretical basis for human behaviour dynamics of comments on Blog posts.

**Key words:** human dynamics; Poisson process; power-law distribution; comments on Blog posts.

[**] Project 70871082 supported by the National Natural Science Foundation of China (Grant No 70871082) and Project S30504 supported by the Shanghai Leading Academic Discipline Project (Grant No S30504).
Corresponding author. *E-mail address*: phd5816@163.com.

# 博客评论的人类行为动力学实证研究和建模[*]

郭进利

（上海理工大学管理学院工业工程研究所，上海，200093）

**摘要**：由于 Internet 上的在线通信信息可以存储在数据库中，它为研究人类动力学提供了机遇。本文基于网上博客评论的数据统计，给出评论时间间隔分布的实证研究。发现人们对于某个话题的兴趣逐渐消失，评论的时间间隔服从幂律分布。根据这个特征，我们提出一个兴趣逐渐消失的人类行为动力学模型。用非齐次 Poisson 过程对模型进行严格分析，获得事件的时间间隔分布的解析表达式。结果表明，这个模型的时间间隔服从幂律分布，幂律指数由两个参数调节，其范围在区间（1，∞）中．理论预测和实证结果相吻合。

**关键词**：人类动力学；Poisson 过程；幂律分布，博客帖子评论。

## 1．引言

人们每天参与大量不同类型的活动,如，发电子邮件,处理销售订单到参加娱乐活动等等。上世纪人们试图用计数过程和排队模型刻画人类行为。在这方面最简单也是应用最广的是 Poisson 过程。20 世纪初，Erlang 在研究电话交换台呼叫过程中进一步讨论了 Poisson 分布，并将 Poisson 分布推广得到了爱尔朗分布。此后，Poisson 过程在物理学、通讯技术、交通运输和管理科学等领域得到了广泛的应用。[1-2]

Poisson 过程特征为事件的到达时间间隔相互独立而且服从同一个指数分布。然而，一般人类活动的时间间隔不一定相互独立，也不一定有相同的分布。2005 年，Barabási 通过对 E-mail 和爱因斯坦等人信件的数据实证发现，人类的许多活动的时间间隔服从幂律分布，这个发现对传统人类活动模型中任务按照 Poisson 过程描述提出了挑战。Barabási 的实证表明

人类活动方式是相当不均匀的，短期高活跃状态分隔长期的休止状态[3,4]。这种不均匀是以给定任务的两次连续执行时间间隔的重尾分布为特征。引起这个特征的现象是多种多样的，人们按照优先级执行任务[3-8]、基于记忆自适应兴趣的驱动[9,10]和对网页浏览的兴趣均都有可能引起重尾分布。Grabowski 在一类社会网络中研究了人们对一件事件感兴趣的时间长度[7]，也发现了人类动力学中有趣的标度律。很多实证结果表明人类行为动力学表现出指数为 1 和 1.5 的幂律分布，以至于 Vázquez 等人在文献[11]中将人类动力学划分为两个适普类。对这两个适普类划分的质疑是周涛等人关于钱学森信件的实证[12]。文献[13]中意识到非齐次 Poisson 过程在人类行为动力学研究中的重要性，形成了用非齐次 Poisson 过程研究人类动力学的思路。最近，文献[14]用级联式的非齐次 Poisson 过程研究人类通信的适普性。其实，人类行为是非常复杂的。人们可能对某件事情随时间推移不断的表现出浓厚兴趣，比如，吸毒和吸烟随着毒瘾越来越大吸食的次数越来越频繁。人们也可能对某件事情随时间推移兴趣越来越淡漠，甚至兴趣消失，比如，央视的综艺大观和实话实说栏目最初收视率非常高，随着时间的推移人们慢慢的对节目内容失去兴趣，现在它们的收视率已经不是很高了。有些事情和个体的差异也有很大的关系，有人对上网聊天兴趣越来越高，而有人对其越来越淡漠。再例如，如果有人在网上发一个关于学术不端的帖子。开始很多人跟贴评论，随着时间推移慢慢没有人关心此事。由此可见，人们对某事件兴趣逐渐消失的变化规律是值得研究的。兴趣消失事件的时间间隔分布是否具有幂律特征？如果具有幂律特征是否属于 Vázquez 等人的适普类？本文的目的是试图回答这些问题。

Internet 的高速发展，为人们相互交流提供了方便平台，也为研究人类行为动力学提供了数据源。在第二节中，我们分别收集网上四个敏感话题的评论数据。通过统计分析发现，虽然这四个博客话题互不相干，但是它们的共同特征是：人们开始对其兴趣极浓，时间长了兴趣消失。实证表明，这些评论的时间间隔分布具有幂律分布特征。在第三节中，根据上一节的实证研究，提出一个兴趣逐渐消失的人类行为动力学模型。在第四节中对我们提出的模型进行分析。分析表明理论预测和实证结果相吻合。第五节是结论和讨论。

## 2. 博客评论时间间隔分布实证研究

我们收集了新浪网博客和中国科学网博客上四个关于科研、教学和研究生培养话题的讨论。通过模拟和利用回归方法对主体数据进行拟合（见图 1）。新浪网一个名人博客 2008 年

10月9日发了一个与诺贝尔奖有关的帖子A，到2010年2月5日收到721条评论。其中，2008年10月份99.2%条评论，2008年11月份0.4%条评论，2008年12月份0.4%条评论，2009年以后没有评论。由此可见，人们开始对这个话题的兴趣极浓，现在已经对它没有什么兴趣。图1(a)是在双对数坐标中评论的时间间隔分布图，图中的直线是利用回归方法得到主体数据的拟合直线，相关系数 R=0.9644。从图 1(a)可见这个帖子的评论时间间隔分布是指数近似为1.9777的幂律分布。

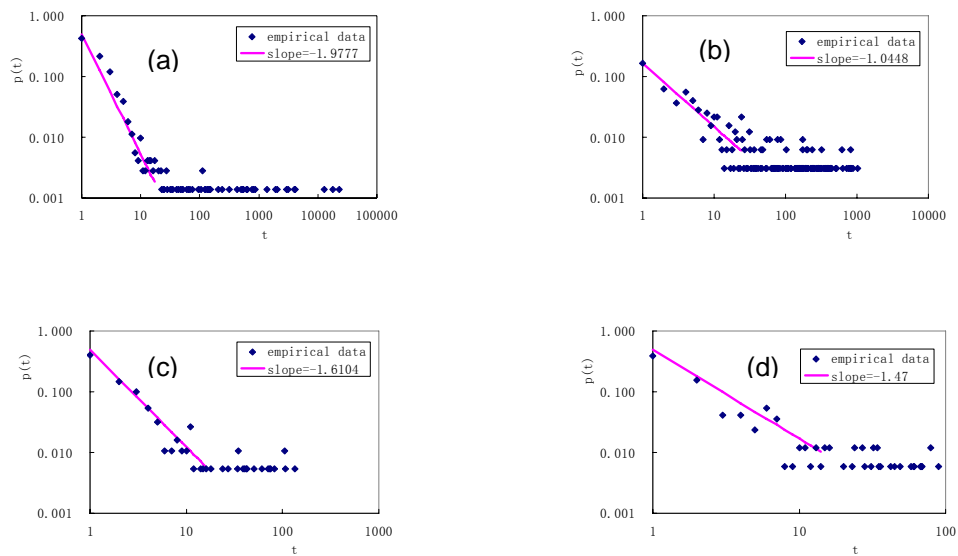

**图 1.** 双对数坐标系中评论时间间隔分布. (a) 帖子 A 的评论时间间隔分布. (b) 帖子 B 的评论时间间隔分布.(c) 帖子 C 的评论时间间隔分布. (d) 帖子 D 的评论时间间隔分布.

中国科学网博客上三个博客分别发表了帖子B、C和D。2009年5月17日发了关于科研办学敏感话题的帖子B，到2010年2月5日收到320条评论。2009年5月份占55.4%，2009年6月份占21%，2009年7月份占14%，2009年8月份占9%，2009年9月份占0%，2009年10月份占0.3%，11月份占0.3%，2009年12月份以后没有评论。可见，人们对这个话题开始非常热衷，慢慢淡漠，最后完全没有了兴趣。图1(b)是在双对数坐标中评论时间间隔分布图，利用回归方法得到主体数据的拟合直线，相关系数 R=0.8564。从图 1(b)可见这个帖子的评论时间间隔分布是指数近似为1.0448的幂律分布。

2009年6月4日发了帖子C。到2010年2月收到了189条评论. 其中，2009年6月占93.66%，2009年7月占3.17%, 2009年10月占0.53%, 2009年12月占2.11%, 2010年1月占0.53%, 2010年2月以后没有评论. 图 1(c)是在双对数坐标中评论时间间隔分布图，利用回归方法得到主体数据的拟合直线，相关系数 R=0.9324。从图 1(c)可见这个帖子的评论时间间隔分布是指

数近似为 1.6104 的幂律分布。

2010 年 1 月 6 日发了贴子 D。到 2010 年 2 月 10 日收到 171 条评论。1 月 6 日至 1 月 10 日占 63.53%，1 月 11 日至 1 月 20 日占 16.47%，1 月 21 日至 1 月 31 日占 13.53%，2 月 1 日至 2 月 10 日占 6.47%。图 1(d)是在双对数坐标中评论时间间隔分布图，利用回归方法得到主体数据的拟合直线，相关系数 R=0.911。从图 1(d)可见这个帖子的评论时间间隔分布是指数近似为 1.47 的幂律分布。

### 3．模型

Barabási 在文献[3]中指出：当个体依据优先级参数选择任务执行时，任务的等待时间分布服从幂律分布。Barabási 模型定义如下：任务列表有 $L$ 项必须做的任务，当一项新任务加入到列表中时，以概率密度函数 $\rho(x)$ 分配给他一个优先权 $x \geq 0$。初始状态 $t$=0 时，有 $L$ 项新任务加入列表。在每个离散时间步 $t$>0，以概率 $p$ 选择列表中有最高优先权的任务接受服务，以概率 $1-p$ 随机选择一项任务接受服务。选择的任务从列表中立即移出，同时有一项新任务加入列表。文献[3]中，Barabási 在 $p \to 1$ 条件下给出了任务在第 $\tau$ 时间步被执行的概率为：

$$P_\tau \approx \tau^{-1} \qquad (1)$$

从上式可见的，对于最高优先权首选规则的模型任务等待时间呈现指数为 1 的幂律分布。

Barabási 模型刻画了一类人们对任务选择的模型。但是，不是每类事情都可以被看作这类任务。我们有这样的经验，当人们对初次接触的新鲜事物兴趣往往很大，随着时间推移人们对其慢慢失去兴趣。上一节中关于博客评论的实证就是这类兴趣消失的例子。从实证可见，评论时间间隔服从幂律分布。我们对这类兴趣消失的人类行为动力学现象刻画为：(a) 在不同的时间间隔事件发生的次数是相互独立的；(b) 事件发生率随时间线性递减，即在时刻 $t$ 事件的发生率为 $\lambda(t) = b/(at+1)$，其中，$a \geq 0, b > 0$；(c) 在 $t$ 与 $t + dt$ 之间事件发生一次的概率为 $\lambda(t)dt$，短时间 $dt$ 内事件几乎不会发生两次以上。

事件的发生率为 $\lambda(t) = b/(at+1)$ 体现出人们对这个事件的兴趣随时间不断衰减，长时间后兴趣完全消失。$a$ 越大兴趣衰减越快，$b$ 越大兴趣衰减越慢。

## 4．模型分析

随机过程 $\{N(t), t \geq 0\}$ 称为一个计数过程，若 $N(t)$ 表示到时刻 $t$ 为止已发生的"事件"的总数，因此，一个计数过程满足，（i） $N(t) \geq 0$；（ii） $N(t)$ 是整数值；（iii） 若 $s<t$，则 $N(s) \leq N(t)$；（iv） 若 $s<t$ 时，$N(t)-N(s)$ 等于区间 $(s,t]$ 中发生的事件个数。如果在不相交的时间区间中发生的事件个数是独立的，则称计数过程有独立增量。若在任意时间区间中发生的事件个数的分布只依赖于时间的区间长度，则称计数过程有平稳增量。

假设 $T_n$ 表示计数过程 $\{N(t), t \geq 0\}$ 第 $n$ 个事件与第 $n-1$ 个事件之间的时间间隔，则过程的点发生时间序列为

$$S_0 = 0, \quad S_n = \sum_{i=1}^{n} T_i, \quad n=1,2,\cdots \qquad (2)$$

计数过程 $\{N(t), t \geq 0\}$ 的强度(若定义的极限存在)

$$\lambda(t) = \lim_{\Delta t \to 0} \frac{P\{N(t+\Delta t) - N(t) > 0\}}{\Delta t} \qquad (3)$$

是一个能很好地刻画过程计数性质的量，可以通过它来描述一个计数过程。$\lambda(t)dt$ 是在时间 $t$ 和 $t+dt$ 之间事件出现的概率。齐次 Poisson 过程作为一类重要的计数过程，它的一个主要特征是具有平稳独立增量。我们将齐次 Poisson 过程的增量平稳性移去，得到非齐次 Poisson 过程[15]。由文献[15]可见，我们的模型为强度为 $\lambda(t)$ 非齐次 Poisson 过程。

**引理 1**[2]. 设 $\{N(t), t \geq 0\}$ 为具有强度函数 $\lambda(t)$ 的非齐次 Poisson 过程，对于任意给定的正整数 $n$ 和 $r$，过程的 $n$ 个点发生时间 $S_{r+1}, S_{r+2}, \cdots, S_{r+n}$ 的联合分布密度函数是

$$f_{S_{r+1},S_{r+2},\cdots,S_{r+n}}(t_{r+1},t_{r+2},\cdots,t_{r+n}) = \begin{cases} (\prod_{i=1}^{n}\lambda(t_{r+i}))\exp\{-\int_{0}^{t_{r+n}}\lambda(x)dx\}\frac{1}{r!}\left(\int_{0}^{t_{r+1}}\lambda(x)dx\right)^r \\ \qquad\qquad\qquad\qquad\qquad\qquad\qquad 0<t_{r+1}\leq\cdots\leq t_{r+n} \\ 0 \qquad\qquad\qquad\qquad\qquad\qquad 其它情形 \end{cases}$$

**引理 2**[2]. 设 $\{N(t), t \geq 0\}$ 为具有强度函数 $\lambda(t)$ 的非齐次 Poisson 过程，对于任意给定的正整数 $n$，过程的 $n$ 个点发生时间为 $S_1, S_2, \cdots, S_n$，第 $n+1$ 个事件与第 $n$ 个事件之间的时间间隔 $T_{n+1}$ 的条件分布函数是

$$P\{T_{n+1} \le t \mid S_n = x_n, \cdots, S_1 = x_1\} = 1 - \exp\{-\int_{x_n}^{x_n+t} \lambda(u)du\}$$

由引理 1 和引理 2，我们有

$$P\{T_{n+1} > t\} = \int \cdots \int P\{T_{n+1} > t \mid S_n = x_n, \cdots, S_1 = x_1\} f_{S_1,\cdots,S_n}(x_1,\cdots,x_n)dx_1 \cdots dx_n$$

$$= \int_0^\infty dx_n \int_0^{x_n} dx_{n-1} \int_0^{x_{n-1}} dx_{n-2} \cdots \int_0^{x_3} dx_2 \int_0^{x_2} \exp\{-\int_0^{x_n+t} \lambda(u)du\} \prod_{i=1}^n \lambda(x_i)dx_1$$

$$= \int_0^\infty \exp\{-\int_0^{x_n+t} \lambda(u)du\} \lambda(x_n)dx_n \int_0^{x_n} \lambda(x_{n-1})dx_{n-1} \int_0^{x_{n-1}} \lambda(x_{n-2})dx_{n-2} \cdots \int_0^{x_3} \lambda(x_2)dx_2 \int_0^{x_2} \lambda(x_1)dx_1$$

$$= \frac{b^n}{(n-1)!a^{n-1}} \int_0^\infty \frac{\ln^{n-1}(ax+1)}{(a(x+t)+1)^{\frac{b}{a}}(ax+1)}dx$$

$$= \frac{b^n}{(n-1)!a^n} \int_0^\infty \frac{y^{n-1}}{(e^y+at)^{\frac{b}{a}}}dy$$

因此，时间间隔 $T_{n+1}$ 的分布函数为

$$F_{T_{n+1}}(t) = 1 - \frac{b^n}{(n-1)!a^n} \int_0^\infty \frac{y^{n-1}}{(e^y+at)^{\frac{b}{a}}}dy, \tag{4}$$

当 $a=0$ 时，时间间隔 $T_{n+1}$ 的分布函数为

$$F(t) = \lim_{a \to 0} F_{T_{n+1}}(t) = 1 - \lim_{a \to 0} \frac{b^n}{(n-1)!a^n} \int_0^\infty \frac{y^{n-1}}{(e^y+at)^{\frac{b}{a}}}dy = 1 - e^{-bt}, \tag{5}$$

式(5)表明，当 $a=0$ 时，时间间隔 $T_{n+1}$ 的分布函数为指数分布，模型退化为齐次 Poisson 过程。

当 $a>0$ 时，由式(4)，我们知道时间间隔 $T_{n+1}$ 分布函数的密度函数为

$$f_{T_{n+1}}(t) = \frac{b^{n+1}}{(n-1)!a^n} \int_0^\infty \frac{e^{-(\frac{b}{a}+1)y} y^{n-1}}{(1+ae^{-y}t)^{\frac{b}{a}+1}}dy, \tag{6}$$

当 $T$ 充分大时，

$$f_{T_{n+1}}(t) \approx \frac{b^{n+1}}{(n-1)!a^n} \int_0^T \frac{e^{-(\frac{b}{a}+1)y} y^{n-1}}{(1+ae^{-y}t)^{\frac{b}{a}+1}}dy, \tag{7}$$

由积分中值定理，存在 $\xi \in [0,T]$，使得，

$$f_{T_{n+1}}(t) \approx \left(\frac{b^{n+1}}{(n-1)!a^n}\int_0^T e^{-(\frac{b}{a}+1)y}y^{n-1}dy\right)\frac{1}{(1+ae^{-\xi}t)^{\frac{b}{a}+1}},$$

$$= \frac{b^{n+1}}{(a+\alpha)^n}\left(1-\frac{1}{(n-1)!}\Gamma(n,(\frac{b}{a}+1)T)\right)\frac{1}{(1+ae^{-\xi}t)^{\frac{b}{a}+1}} \tag{8}$$

这里，$\Gamma(n,x)=\int_x^\infty e^{-y}y^{n-1}dy$ 是不完全的 Gamma 函数。

由于 $1-F(t)$ 为时间间隔的累积概率分布，因此，从(8)可见，时间间隔 $T_{n+1}$ 的分布函数是幂律指数为 $\gamma = \frac{b}{a}+1$ 的幂律分布。由于 $a>0, b>0$，故指数范围区间 $(1,\infty)$ 内。实证得到的结果：$\gamma = 1.997, 1.0448, 1.6104, 1.47$. 因此，分析表明理论预测和实证结果相吻合。

## 5．结论与讨论

我们通过对网上贴子评论时间间隔的实证研究，发现人们对新帖子兴趣较大，随着时间的推移人们对其兴趣衰减，以至于可能失去兴趣。根据这个特征建立兴趣消失的人类行为动力学模型。通过对模型的严格分析，发现兴趣消失的人类行为动力学的事件间隔时间服从幂律分布，幂律指数可以是任何大于 1 的实数。我们的模型为实证结果：$\gamma = 1.997, 1.0448, 1.6104, 1.47$ 提供了理论依据。体现出了人类动力学中幂律指数的多样性，说明人类行为动力学机制不像文献[11]所描述的那样简单，预示着人类行为动力学机制的复杂性。

文献[9]中研究了兴趣驱动的人类动力学模型，这是一个时间离散模型。作者在模型中假设衰减函数为 $\frac{1}{1+ai}, i=1,2,\cdots$。文献[9]中指出：若事件在时刻 $t$ 发生，则在时刻 $t+1$ 发生事件的概率为 $\frac{1}{1+a}$. 从文献[9]中的式(1)可知，若事件在时刻 $t$ 没有发生，则在时刻 $t+1$ 发生事件的概率不小于 $\frac{1}{1+a(t+1)}$. 则

P{在时刻 $t+1$ 事件发生}

= P{在时刻 $t+1$ 事件发生／在时刻 $t$ 事件发生}+

P{在时刻 $t+1$ 事件发生／在时刻 $t$ 事件不发生}

$\geq \frac{1}{1+a}+\frac{1}{1+a(t+1)}$

如果将文献[9]中模型的衰减函数扩展为 $\frac{b}{1+ai}, i=1,2,\cdots$，则要求满足条件

$$\frac{b}{1+a} < 1. \qquad (9)$$

文献[9]中提出的模型是时间离散模型。从文献[9]中的分析可知事件是等时间间隔发生，这与博客评论的时间间隔不相符。本文提出的是时间连续模型，假设发生率（数学上也称强度函数）为 $\lambda(t) = \frac{b}{at+1}$. 对于充分小的 $dt$，在时间 $t$ 和 $t+dt$ 之间事件出现的概率是 $\lambda(t)dt$。

因此，发生率 $\lambda(t)$ 与文献[9]中的衰减函数不完全相同，也不需要条件（9）的约束。更重要的是，本文对这类问题提供了一个更严密的研究思路和估计事件发生时间间隔分布的途径。

**致谢**